\documentclass{emulateapj}
\usepackage{lscape}
\input{epsf.sty}

\newcommand{\solar}{\ifmmode_{\mathord\odot}\else$_{\mathord\odot}$\fi} 
\newcommand{\dgr}{\ifmmode {^\circ}\else $^\circ $\fi}  
\newcommand{\arcs}{\ifmmode {'' }\else $'' $\fi}  
\newcommand{\arcm}{\ifmmode {' }\else $' $\fi}    
\newcommand{\mstar}{\ifmmode {M_{HI_\ast}}\else $M_{HI_\ast}$\fi}
\newcommand{\msolar}{\ifmmode {M_\odot} \else M$_\odot$\fi}
\newcommand{\gapprox}{\ifmmode \buildrel > \over {_\sim} \else $\buildrel >\over {_\sim}$\fi}
\newcommand{\lapprox}{\ifmmode \buildrel < \over {_\sim} \else $\buildrel <\over {_\sim}$\fi}

\begin{document}

\shortauthors{Rosenberg, Ashby, Salzer, \& Huang}

\title{The Diverse Infrared Properties of a Complete Sample of Star-Forming
Dwarf Galaxies}
\shorttitle{Infrared Properties of Star-Forming Dwarfs}

\author{Jessica L. Rosenberg\footnote{National Science Foundation Astronomy and 
Astrophysics Postdoctoral Fellow}, Matthew L. N. Ashby}
\affil{Harvard-Smithsonian Center for Astrophysics, 60 Garden Street MS 65, 
Cambridge, MA 02138}
\email{jlrosenberg@cfa.harvard.edu}
\author{John J. Salzer} 
\affil{Wesleyan University, Department of Astronomy, Middletown, CT 06459}
\author{Jia-Sheng Huang}
\affil{Harvard-Smithsonian Center for Astrophysics, 60 Garden Street MS 65, 
Cambridge, MA 02138}

\begin{abstract}

We present mid-infrared Spitzer Space Telescope observations of a 
complete sample of star-forming dwarf galaxies selected from the KPNO
International Spectroscopic Survey. The galaxies span a wide range in 
mid-infrared properties. Contrary to expectations, some of the galaxies 
emit strongly at 8 $\mu$m indicating the presence of hot dust and/or 
PAHs. The ratio of this mid-infrared dust emission to the stellar emission is
compared with the galaxies' luminosity, star-formation rate, metallicity, and
optical reddening. We find that the strength of the 8.0$\mu$m dust emission to 
the stellar emission ratio is more strongly correlated with the star-formation 
rate than it is with the metallicity or the optical reddening in these systems. 
Nonetheless, there is a correlation between the 8.0
$\mu$m luminosity and metallicity. The slope of this luminosity-metallicity 
correlation is shallower than corresponding ones in the B-band and 3.6 $\mu$m.
The precise nature of the 8.0$\mu$m emission seen in these galaxies (i.e., PAH
versus hot dust or some combination of the two) will require future study,
including deep mid-IR spectroscopy.

\end{abstract}


\section{Introduction}

Low-mass, low-metallicity systems are the building blocks in hierarchical galaxy
formation scenarios. Star-formation processes in these
systems may play a critical role in the evolution of galaxies including the
feedback of metals into the intergalactic medium \citep{maclow1999,ferrara2000,
silich2001}, the suppression of low-mass halos that are over-produced in 
semi-analytic models of galaxy formation \citep{kravtsov2004}, and the color 
evolution of galaxies. Nearby actively 
star-forming dwarf galaxies are possible analogs to these high redshift systems.
These galaxies are gas-rich, metal poor, and undergoing bursts of star formation 
in the local universe. As low-redshift systems, these galaxies can be studied in
much greater detail than is possible for galaxies in the high redshift universe.

At visible wavelengths dwarf star-forming galaxies are very blue and exhibit 
strong nebular
emission lines. Many of the systems are very compact with morphologies like
large HII regions, although some do resemble dwarf irregular or dwarf spiral 
galaxies. At longer wavelengths dust emission from reprocessed starlight is a 
good tracer of star formation in galaxies \citep{dwek1998}. However, observations 
of a sample of emission-line galaxies with the Infrared Astronomical Satellite (IRAS) 
have indicated that dwarf star-forming galaxies are less likely to be detected 
in the far infrared than their higher luminosity counterparts \citep{salzer1988}.
These galaxies could be more difficult to detect in the FIR due to a lack of dust, 
a different dust grain size distribution, or a different dust temperature distribution. 
The mid-infrared to millimeter spectral energy distributions (SEDs) of a small number
of nearby dwarf galaxies selected for their high star-formation rates indicate 
that some dwarf star-forming systems do contain a significant amount of dust but 
that the systems probably lack PAHs and have a small average dust grain size 
\citep{galliano2005,galliano2003}. 

The launch of the Spitzer Space Telescope \citep{werner2004} has opened up a new 
window on the study of star-formation processes in galaxies. The IRAC instrument
\citep{fazio2004} in particular allows us to measure both the stellar emission
and the hot dust and PAH emission in local galaxies. One of the first dwarf 
galaxies to be examined in detail with Spitzer was the blue compact dwarf galaxy 
SBS 0335-052 \citep{houck2004}. This galaxy, with one of the lowest known 
metallicities \citetext{Z$\approx$ Z$_\odot$/20, \citealp{izotov1997}}, is known to 
have dust
but it lacks PAH emission \citep{thuan1999, plante2002}. The mid-infrared flux
from the galaxy is dominated by a cold dusty envelope ($\sim$65 K) and shows 
evidence for a warm dust component ($\sim$150 K). While SBS 0335-052 is an
unusual system, \citet{hogg2005} examined the mid-infrared colors of $\sim$10 low 
luminosity galaxies and \citet{engelbracht2005} examined a sample of low 
metallicity galaxies with Spitzer and both concluded that a significant fraction
of the low-luminosity and low-metallicity systems show a deficit of PAH emission.
In the case of the \citet{hogg2005} data, there are no
spectra to discern the difference between PAHs and a hot dust continuum but
\citet{engelbracht2005} do have spectra and draw the same conclusion -- the 
mid-infrared shows a PAH deficit but an 8.0 $\mu$m excess 
consistent with hot dust emission.

Overall, only a small sample of dwarf galaxies has been looked at with Spitzer
to date. We present the first detailed look at a statistically complete sample of 
dwarf galaxies known to be undergoing a burst of star formation. In \S 2 we
discuss the selection of this sample, the optical and infrared observations and
the data reduction. In \S 3 we present the observational results for the sample
from the optical, near infrared, and mid-infrared data and then we discuss
those results in \S 4. 

\section{Sample Selection, Observations, and Data Reduction}

\subsection{KISS Observations}

The KPNO International Spectroscopic Survey (KISS) is a 
modern objective-prism survey that combines the methodology of many of the 
classic wide-field color- and line-selected surveys \citetext{e.g., 
\citealp{markarian1967, Smith1976, macalpine1977, wasilewski1983, 
zamorano1994}} with the higher
sensitivity of a CCD detector.  The survey method is described in detail by 
\citet{salzer2000}.  KISS selects objects for inclusion in the survey lists if 
they possess a strong ($>$ 5$\sigma$) emission line in their low-dispersion 
objective-prism spectra.  The survey has been carried out in two distinct 
spectral regions: the blue portion (4800 -- 5500 \AA) where the primary line 
observed is [\ion{O}{3}]$\lambda$5007, and the red region (6400 -- 7200 \AA) where 
galaxies are selected by their H$\alpha$ emission.  To date, the survey has 
produced one list of emission-line galaxy (ELG) candidates in the blue 
\citep{salzer2002} and three in the red \citetext{\citealp{salzer2001, gronwall2004, 
jangren2005}, hereafter KR3}. For the current study, we have selected a sample 
of dwarf star-forming galaxies from KR3. This survey list was derived from 
objective-prism data taken of the NOAO Deep Wide-Field Survey (NDWFS) area 
located in Bo\"otes.  Both of the NDWFS fields were targeted by KISS to take 
advantage of the tremendous amount of multi-wavelength data planned for these 
areas.  

All of the KISS ELG candidates in the Bo\"otes field possess follow-up slit spectra
\citep{salzer2005b}. Higher dispersion follow-up spectra are necessary in order 
to verify the reality of the putative emission lines seen in the objective-prism 
spectra. In addition, the survey data alone do not have sufficient dispersion or 
spectral range to provide accurate redshifts or to distinguish between the 
various activity types that might be present in a line-selected sample (e.g.,
star-forming galaxies vs.~AGNs). These follow-up spectra provide us with a 
great deal of useful information (e.g., accurate redshifts, emission-line fluxes 
and line ratios, reddening and metallicity estimates). The combination of the 
accurate B and V photometry from the original survey lists with these follow-up 
spectra allow for the construction of a fairly complete picture of the properties of the 
KISS ELGs.

The selection of the KISS galaxies from the KR3 list was defined by several
criteria. For emission-line galaxies selected by their H$\alpha$ emission (in 
the red), the redshift limit of the survey is $z = 0.095$. The two selection 
criteria used to define the sample were that the galaxies exhibit spectra 
consistent with excitation by star-formation processes (i.e., AGNs were 
excluded), and that they have a B-band absolute magnitude M$_B > -$18.0 (for 
H$_0$ = 75 km s$^{-1}$ Mpc$^{-1}$). Thus these galaxies are selected to be 
star-forming dwarf galaxies. These criteria produced a list of 26 galaxies 
within the NDWFS Bo\"otes area. However, of those 26 galaxies, only 19 overlap
with the Spitzer Shallow Survey area because the Spitzer survey field is 
smaller than the NDWFS field. We discuss only these 19 galaxies in our analysis.
Table 1 presents the KISS data for this sample.

All magnitudes discussed in this paper are Vega relative magnitudes.

\subsection{2MASS and NOAO Deep Wide-Field Survey Data} 							 
														 
The field from which our sample of 19 dwarf star-forming galaxies was drawn has					 
multi-wavelength optical and infrared coverage from the NDWFS 							 
\citep{jannuzi1999,jannuzi2005,dey2005} and the 2 Micron All-Sky Survey (2MASS). 
The NDWFS provides B, R, and I-band data that complement the broadband B and V 
band optical data from the KISS survey. K-band data from the NDWFS are also 
available for 11 of the 19 galaxies in the field. The K-band for the remaining
galaxies are not yet publicly available from the NDWFS.  

The 2MASS data have been included in this table for completeness and as a
reference for future work with this sample even though we
do not use the data here. The 2MASS data are only deep enough to detect four 
galaxies in this sample.  Because these
four sources are compact and near the 2MASS detection limits, they are only
identified in the point source catalog. The photometry available from the NDWFS
is more reliable than the 2MASS data. Note that for the 2 sources for which
2MASS and NDWFS data exist, the NDWFS measurements are much  brighter. The 2MASS
point source measurements are missing a lot of the extended galaxy flux that is
below the detection limit of the survey.

The agreement between the KISS and NDWFS B-band photometry is excellent for all 
of the galaxies except KISS 2344.  This object is very extended and was not well 
fit by the automated NDWFS software (SExtractor ``auto" magnitudes are used,
\citealp{bertin1996}), which underestimated the 
flux of this source by a magnitude relative to the KISS value.  For the rest
of the sample, however, the average difference between the KISS and NDWFS 
magnitudes for the rest of the sample is only 0.04 magnitudes. 
Table 2 contains the NDWFS and 2MASS photometry for the 19 galaxies in the
sample. 

\subsection {Spitzer Observations}

\begin{figure}
\caption{Three color composite images at 3.6, 4.5, and 8.0 $\mu$m (blue, green,
and red respectively). These images have not been point-source subtracted. 
Redshifts are indicated for all objects. JPEG file of figure is included with 
the download.\\ \\ \\ \\ }
\label{fig:images}
\end{figure}

Using the Infrared Array Camera (IRAC) aboard the Spitzer Space Telescope,
the majority of the NDWFS was mapped at 3.6 - 8.0 $\mu$m in January 2004
\citep{eisenhardt2004}. This project, known as the IRAC Shallow Survey, 
covered 8.5 square degrees -- most but not all of the NDWFS field. The observed
region was selected to have the lowest background in the IRAC bands. Because 
not all of the NDWFS field is covered, three of the 26 KISS galaxies in this 
field are just off the eastern edge of the field and 
four are in the southeast corner of the field where there is a hole in the 
coverage. For the remainder of the paper we discuss only the 19 sources with 
coverage in the Spitzer IRAC bands. Each position in the survey field was 
covered with three 30 s IRAC frames, resulting in a depth of 19.1, 18.3, 
15.9, and 15.2 Vega magnitudes ($5\sigma$) at 3.6, 4.5, 5.8, and 8.0$\mu$m, 
respectively. This depth was reached by tiling the 5\arcmin\ $\times$ 5\arcmin\ 
IRAC FOV over the field. 

The basic image processing and mosaicing of the Spitzer data are discussed in 
detail by \citet{eisenhardt2004}. From the resulting mosaiced images (Version 1.1), postage 
stamps 100 $\times$ 100 pixels in size were extracted for our sample galaxies 
(except for KISS 2344 which is more extended so we extracted a 200 $\times$ 200 
pixel postage stamp, the pixel scale for all postage stamps is 0.86\arcs\ per 
pixel). 

Figure \ref{fig:images} shows three color images for all of the galaxies in the
sample. The images, created prior to the
point source subtraction, provide a look at the colors and morphologies of
the objects in the sample.

Because of the high point source density in the fields, we subtracted point
sources from the 3.6 and 4.5 $\mu$m data prior to deriving photometry for the
galaxies. In the 3.6 $\mu$m band 7 -- 39\% of the flux in the aperture is due
to the point sources in the vicinity of our galaxy. The flux due to point
sources in the 4.5 $\mu$m band is less than in the 3.6 $\mu$m band for most
sources, but the range is from $\sim$3 -- 36\%. The point source removal was 
done using the APEX program within the 
MOPEX package provided by the Spitzer Science Center. The PSF used was
constructed by Marengo et al. (2005) and was smoothed to match the resolution 
of the mosaiced images and then used by the APEX program to identify candidate
point sources in the images. We defined the APEX detection parameters so that
all of the stars were selected, but the result was that non-stellar peaks in
faint parts of the galaxy were also identified. The point sources selected by
APEX were then checked by hand in order to eliminate non-stellar sources from
the point source list. The resulting point sources were then subtracted from the
3.6 and 4.5 $\mu$m images. Overall the point source subtraction worked well, but
the centers of the brightest stars which are non-linear or saturated left a
residual. For the 5.8 and 8.0 $\mu$m data, the point source 
density was not high enough to warrant point source subtraction as was 
done in the shorter wavelength bands. 

\begin{figure*}
\plotone{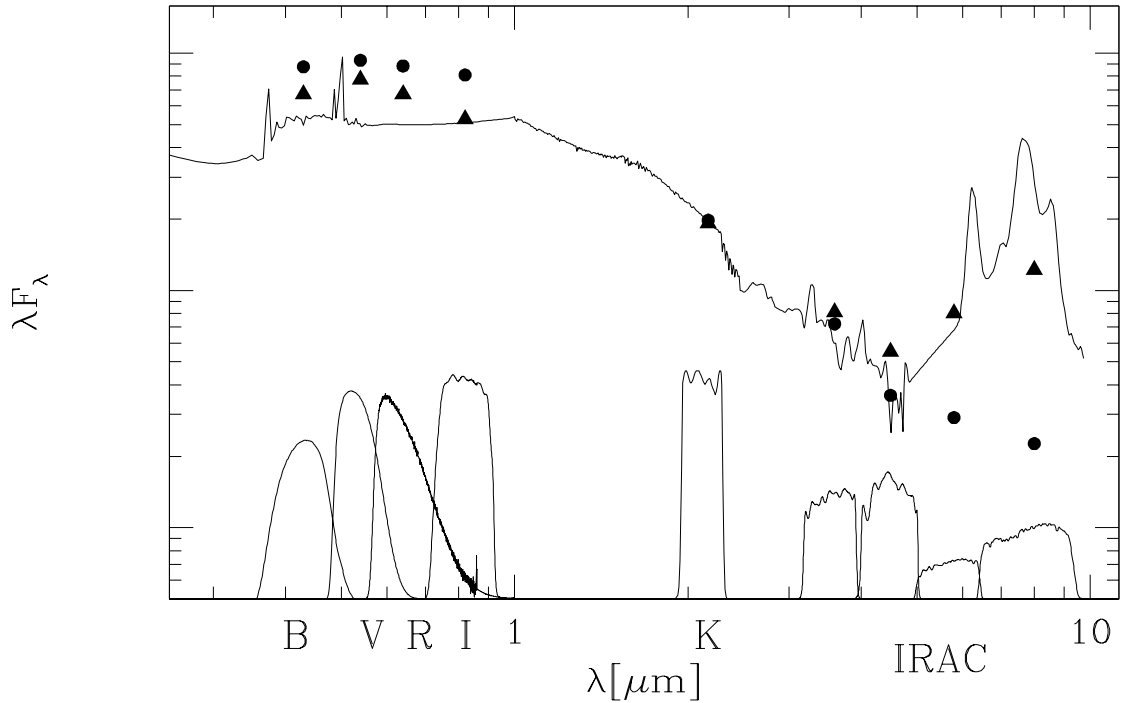}
\caption{An empirical irregular galaxy spectral template showing stellar emission 
at short wavelengths and the dust-emission in the mid-infrared. The template is
based on the \citet{coleman1980} SED in the optical and the \citet{lu2003} SED
in the infrared. The IRAC filter sensitivities as well as B, V, R, I, and K-band 
filter response functions are shown (B and V are Schmidt filters while R and I
are Mosaic filters from http://www.noao.edu/kpno/filters/filters.html. The
K-band data are from http://www.noao.edu/kpno/manuals/onis/mopopulation.html and
the IRAC data are from http://ssc.spitzer.caltech.edu/documents/som/.) The points 
show the SEDs for KISS 2309 
(circles) and KISS 2349 (triangles). These galaxy SEDs have been scaled to the 
same value in the K-band for comparison. These SEDs illustrate the range of 
mid-infrared colors observed in this sample.}
\label{fig:schematic}
\end{figure*}

Galaxy photometry was performed on the processed, mosaiced, and, for the 3.6 
and 4.5 $\mu$m images, point source-subtracted images using the ELLIPSE routine 
within IRAF. Point sources in the 5.8 and 8.0 $\mu$m bands and post point source
subtraction residuals
in the 3.6 and 4.5 $\mu$m bands were masked using ELLIPSE. An elliptical 
aperture was fit to the galaxy, the aperture ellipticity and position angle
were fixed, and then apertures were computed over the background region beyond
the extent of the galaxy. The level determined from the background annulus was
then used to subtract the local background from the image. After
the subtraction, the fitting was rerun using these same aperture parameters out
to the largest radius before the background was reached as determined from the
curve of growth of the flux. The aperture defined 
by the 3.6 $\mu$m data was then used to determine the
flux in the other bands. The results of the IRAC photometry are presented in
Table 3. 

\section {RESULTS}

The combination of the optical and IRAC observations provide a probe of stellar
and dust emission in galaxies in the local universe. Figure \ref{fig:schematic}
shows a template SED for an irregular galaxy. 
This SED template is based on
the \citet{coleman1980} distribution derived in the optical and the \citet{lu2003} 
distribution in the infrared. The template is an empirical derivation of the SED 
extrapolated between the available bands. The optical bands, as well as the 3.6 
and 4.5 $\mu$m IRAC bands probe the stellar emission from the galaxy. The 
3.6 and 4.5 $\mu$m bands, while dominated by the stellar emission, can also be 
influenced by hot dust emission \citetext{particularly the 4.5 $\mu$m band; 
\citealp{roussel2005}}. The 5.8 and 8.0 $\mu$m IRAC bands probe the
Rayleigh-Jeans tail of the hot dust continuum as well as the PAH emission
features from the galaxy. There is a degeneracy between the slope of the hot
dust continuum and the strength of the PAH emission features which does not 
allow us to distinguish between them using broad-band color information. For this 
analysis we focus on the 3.6 $\mu$m and the 8.0 $\mu$m emission as probes of 
the stellar and PAH or hot dust emission in the galaxies that we are studying. 
The [3.6] $-$ [8.0] color provides a measure of the dust-to-stars ratio in the
galaxies.

The points in Figure \ref{fig:schematic} show SEDs for two of the galaxies in this 
sample to illustrate the differences in their properties. The circles show the SED 
for KISS 2309 while the triangles show the SED for KISS 2349. The two SEDs have 
been normalized so that they overlap in the K-band. The optical/near-infrared
photometry has been computed in different apertures for the KISS and NDWFS data,
but the extremely good correlation between the magnitudes measured in the
B-band ($\sigma=0.04$) indicates that the values are consistent. A fixed aperture
is used for all of the mid-infrared measurements, but this are not the same
aperture as used in the optical/near-infrared. Because we use a large aperture 
intended to contain all of the light, the aperture corrections between the 
mid-infrared and the optical/near-infrared should be small. The application of
an aperture correction would result in a small vertical shift for all of the 
mid-infrared points in Figure \ref{fig:schematic}.

\begin{figure*}
\plotone{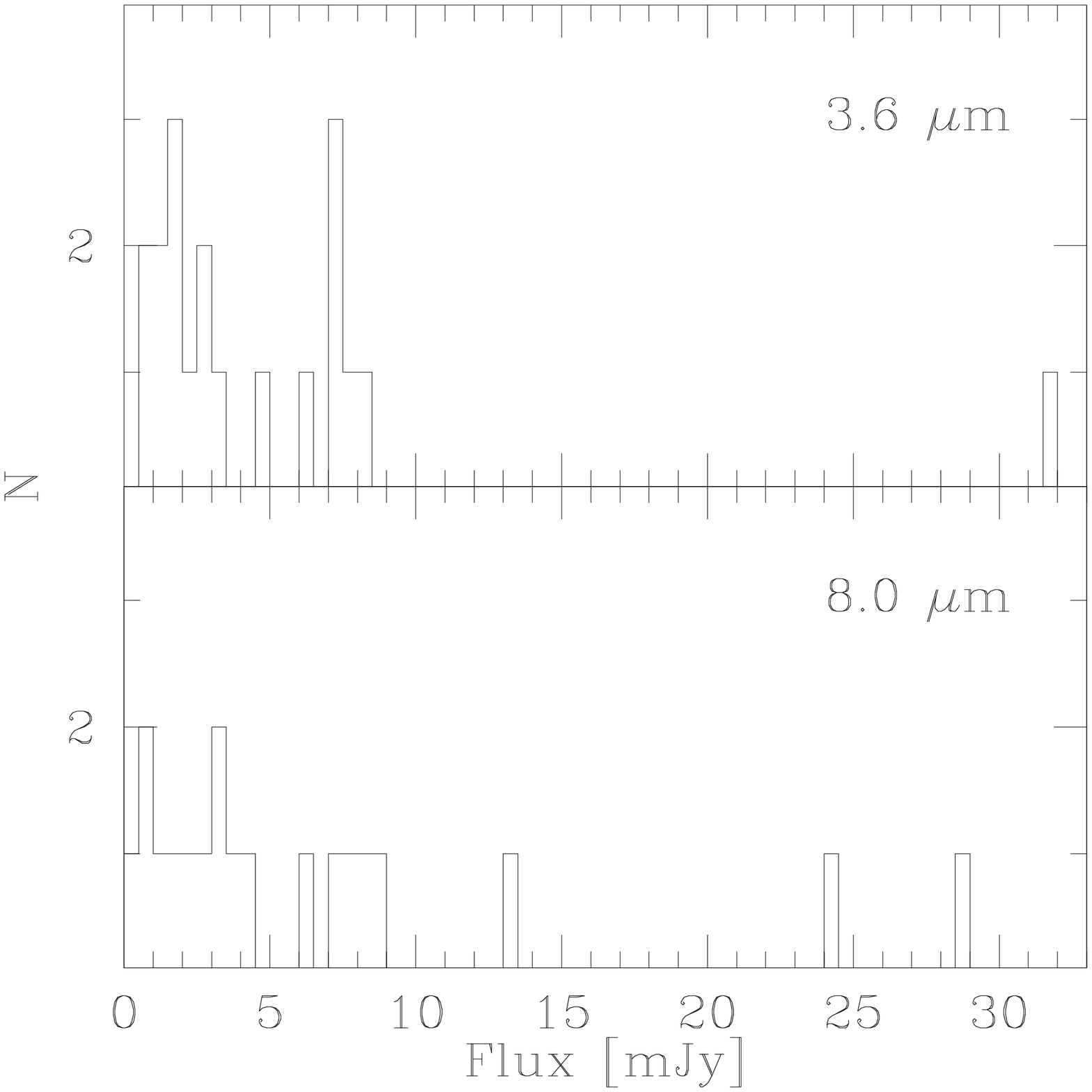}
\caption{The distribution of fluxes (in mJy) at 3.6 and 8.0 $\mu$m for the
sample galaxies.}
\label{fig:fluxhist}
\end{figure*}

Figure \ref{fig:schematic} illustrates the relationship between the mid-infrared 
colors of two sample galaxies and the general shapes of their SEDs. We leave a 
discussion of the detailed SED fitting to a forthcoming paper.
The KISS 2349 SED has a very red [3.6] $-$ [8.0] color ([3.6] $-$ 
[8.0] $= 2.92$). This SED clearly shows evidence for hot dust or, possibly, 
PAH emission. By contrast, the SED of KISS 2309 continues to fall into the 
mid-infrared showing much less hot dust or PAH emission ([3.6] $-$ [8.0] $= 
1.21$), but the colors are still in excess of the extrapolation of the stellar
portion of the template. All of the galaxies in this sample show evidence for
8.0 $\mu$m emission in excess of the stellar light. Both galaxies show 
extremely blue optical colors even 
relative to the K-band flux. These galaxies are dominated by a young stellar 
population even relative to an average irregular galaxy.

\begin{figure*}
\plotone{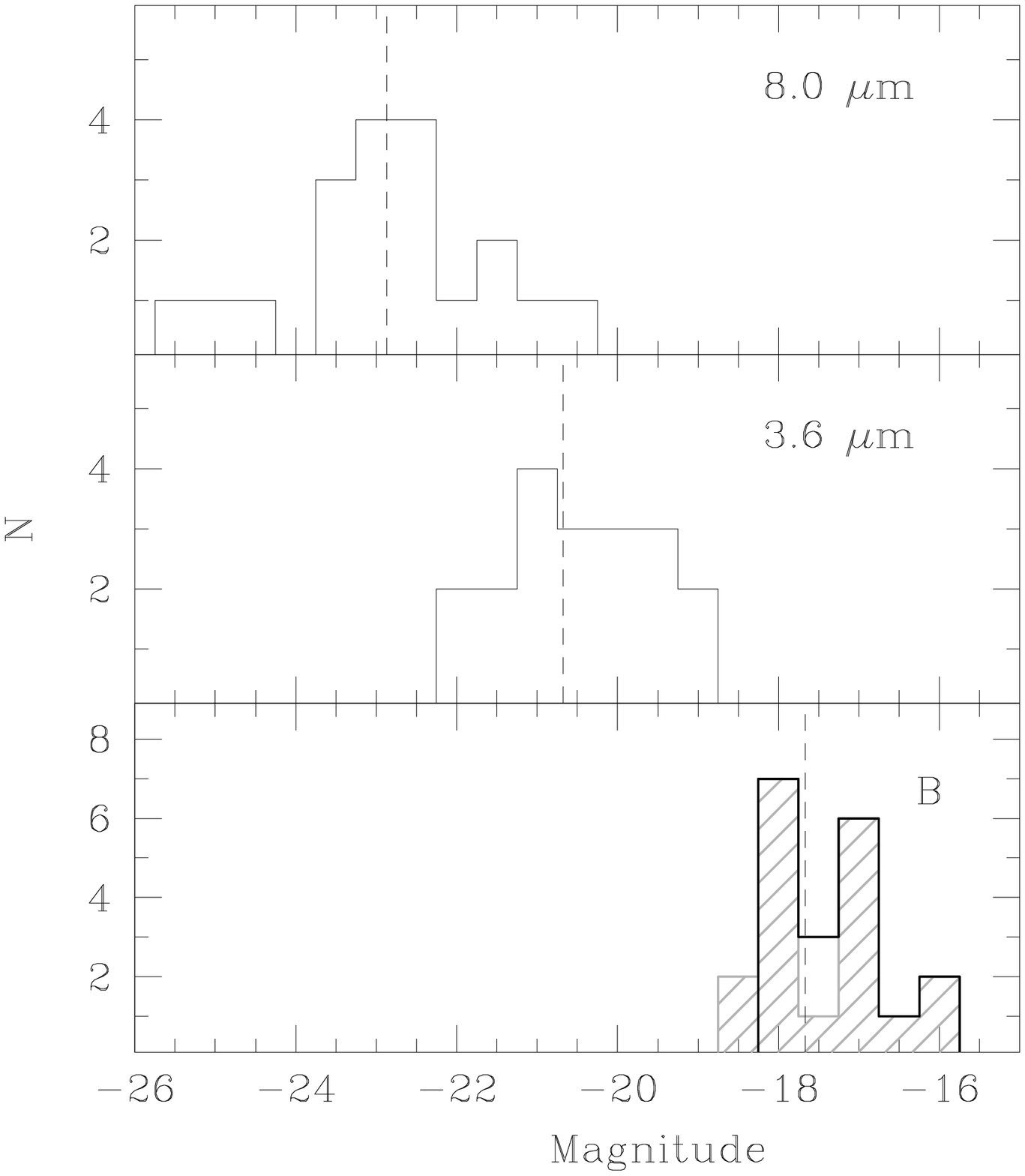}
\caption{Histograms of galaxy absolute magnitudes in the B, 3.6 $\mu$m, and 8.0 
$\mu$m bands. For the B-band histogram, the histogram under the black line indicates
the distrution before correcting for 
internal extinction, while the shaded grey histogram is for the extinction-corrected
distribution. The dashed 
lines indicate the median magnitudes of the sample in each band. Note that 
these galaxies have been selected to have M$_B >$ -18, but there are 3 galaxies 
that are brighter than that limit.}
\label{fig:maghist}
\end{figure*}

\begin{figure}
\plotone{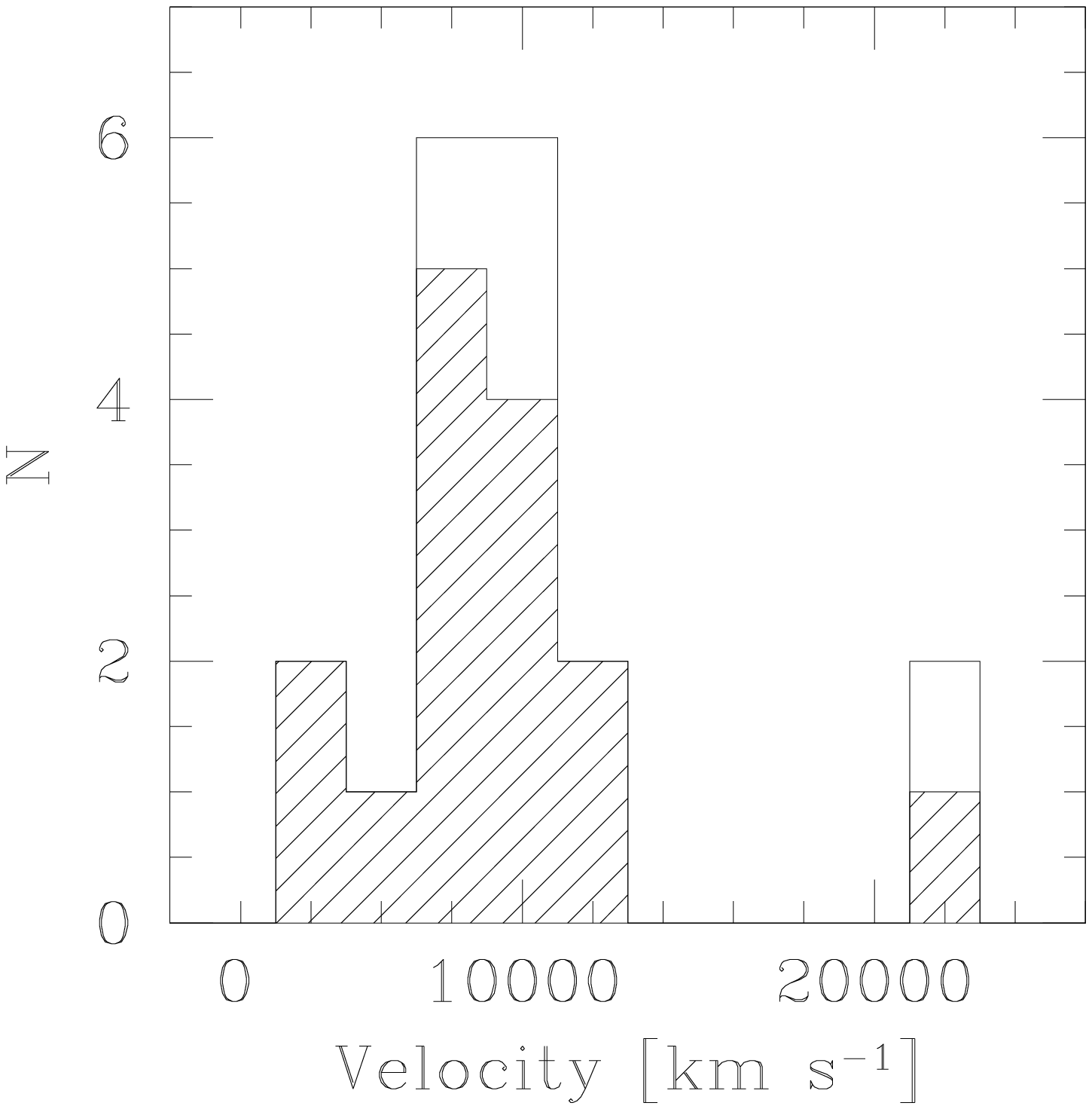}
\caption{The distribution of velocities for the galaxies. The median distance at 
which the galaxies are detected is 115 $h_{75}^{-1}$Mpc for the dwarfs, galaxies
with M$_B < -18.0$,
shown with the shaded histogram and 120 $h_{75}^{-1}$Mpc for the full sample,
which includes galaxies found to be brighter than M$_B = -18.0$ after correcting
for extinction, shown with the open histogram.}
\label{fig:velhist}
\end{figure}

\subsection {Properties of the Sample}

Despite the low detection rate for star-forming dwarf galaxies at
longer infrared wavelengths \citep{salzer1988}, we detected all 19
galaxies in all four bands with only a short exposure time. Only at 5.8 $\mu$m 
are a couple of the galaxies approaching the detection limit of the instrument 
because of the combination of lower sensitivity at these wavelengths and a 
minimum in the spectral energy distribution as can be seen in Figure 
\ref{fig:schematic}. In particular, the detection of these galaxies at 8.0 $\mu$m
indicates more emission from hot dust and/or PAHs than might be expected for
such low-luminosity and generally low-metallicity systems. Figure 
\ref{fig:fluxhist} shows the distribution of fluxes 
at 3.6 and 8.0 $\mu$m for the galaxies in the sample. 

The galaxies have been selected to be low-luminosity systems with a luminosity 
limit of M$_B > -18.0$ -- one galaxy that is at this limit within the errors 
(M$_B = -18.01$) is also included. After
correcting for internal extinction using the ad hoc method developed by
\citet{melbourne2002} we have found that three additional objects are
significantly higher luminosity galaxies. There are two additional galaxies for 
which the spectra are not of high enough quality to determine whether an
extinction correction is appropriate. These are the two systems in Table 1 that
do not have measured metallicities. We continue to 
include all nineteen galaxies in the sample and indicate which galaxies do not 
have a measured internal extinction correction and which galaxies do not have
measured metallicities. We analyse the higher-luminosity galaxies separately 
since they are not actually part 
of the dwarf galaxy sample. The distributions of B, 
3.6 $\mu$m, and 8.0 $\mu$m absolute magnitudes are shown in Figure \ref{fig:maghist}. The 
black B-band histogram shows the uncorrected magnitudes while the shaded grey histogram 
shows the magnitudes corrected for internal extinction. The median magnitudes for 
the sample are M$_{B0}$=-17.67 (corrected for internal extinction), 
M$_{3.6}$= -20.68, and M$_{8.0}$=-22.87, shown in the figure with dashed lines. 

As an objective prism survey, KISS detects galaxies based on the presence of
emission lines rather than a broad-band flux limit. The emission-line detection
makes this survey sensitive to more distant dwarf galaxies than a traditional
flux-limited sample. The dwarf galaxies represented in this study have distances 
up to 295 $h_{75}^{-1}$ Mpc and have a median distance of 115 $h_{75}^{-1}$ Mpc. 
The distribution of velocities for the sample is shown in Figure \ref{fig:velhist}.
The dwarf sample is shown with the shaded histogram, the full sample with the
open histogram.
Despite the large distances of the galaxies and short exposure times for the
Shallow Survey data, all systems were detected. The median absolute 
magnitudes in the Spitzer bands allow us to estimate the maximum distance at 
which these objects could be detected in some of the deeper Spitzer
fields.  In the Chandra Deep Field South and the Hubble Deep Field North (each 
is a 1\dgr\ $\times$ 0.5\dgr\ field) which have 
been observed for 500 seconds, a galaxy with an 8.0 $\mu$m magnitude equal to 
the median for the sample would be a 10$\sigma$ detection at a distance 
of 492 $h_{75}^{-1}$ Mpc. For a much smaller, but deeper survey like the Groth 
Strip (2\dgr $\times$ 10\arcm) which was observed for 3 hours at each pointing, these 
galaxies would be 10$\sigma$ detections at 1060 $h_{75}^{-1}$ Mpc ($z = 0.23$). 

\begin{figure*}
\plotone{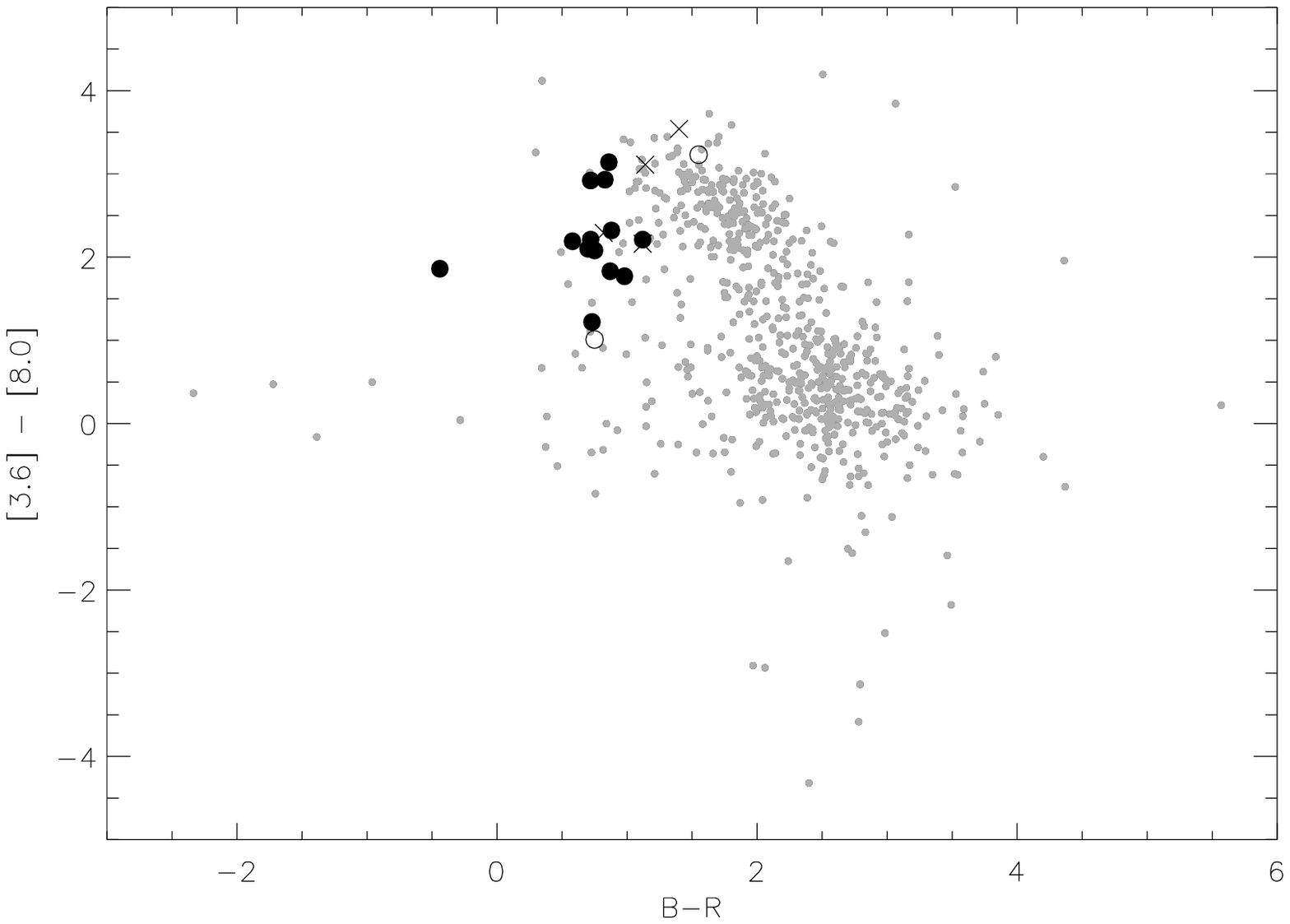}
\caption{The mid-infrared versus optical colors for galaxies in the Shallow
Survey (\citealp{eisenhardt2004}, gray dots) and for KISS dwarf galaxies.
The filled black circles show the true dwarfs in the KISS sample, the $\times$s 
are galaxies that are brighter than M$_B = -18$, and the open circles are the 
galaxies for which an internal extinction correction is not determined. No 
extinction correction has been applied to the optical colors of the KISS 
galaxies for comparison with the values measured for galaxies in the full 
Shallow Survey sample.}
\label{fig:colorplot}
\end{figure*}

The KISS galaxies in this sample are star-forming systems
and, therefore, tend to have blue optical colors. Figure \ref{fig:colorplot}
shows the relationship between the mid-infrared [3.6] $-$ [8.0] color and
the optical B-R color. The small grey dots represent all of the objects
identified as galaxies in the Shallow Survey catalog (version 1.1). For these
objects we use the cataloged magnitudes (``auto" aperture magnitudes determined using 
SExtractor \citep{bertin1996}). As noted by \citet{hogg2005}, the majority of the Shallow 
Survey galaxies fall in two limited regions of color-space. The locus of points with 
[3.6] $-$ [8.0] $\sim 0.5$ contains mostly early-type systems while the 
group of galaxies with redder [3.6] $-$ [8.0] colors are generally 
late-type systems.  The outliers ($0 <$ B-R $<4$ and [3.6] $-$ [8.0] $< -1$) 
in this plot may be objects with poorly
measured magnitudes because of the automated nature of the measurement or they
may be objects with extreme colors. Nevertheless, the plot provides a good way 
to see where the KISS galaxies fall with respect to colors of the galaxies 
within the Shallow Survey field. The KISS galaxies are denoted by three
different point types in this and all subsequent plots. We denote the four 
galaxies that turn out to be more luminous than M$_B = -18$ after the correction 
for internal extinction by $\times$'s. The two galaxies for which the spectra
are not good enough to determine whether a correction for internal extinction
should be applied (or to measure an abundance) are denoted by open circles. The
rest of the galaxies are denoted by filled black circles. These objects have 
enough data to indicate that they are probably true dwarf galaxies by our 
luminosity criteria. The optical colors for the KISS galaxies in Figure
\ref{fig:colorplot} have not been corrected for extinction for consistency with
the colors of the full galaxy sample. While the KISS galaxies are blue optically, 
they span the red end of the range of mid-infrared colors.

\begin{figure}
\plotone{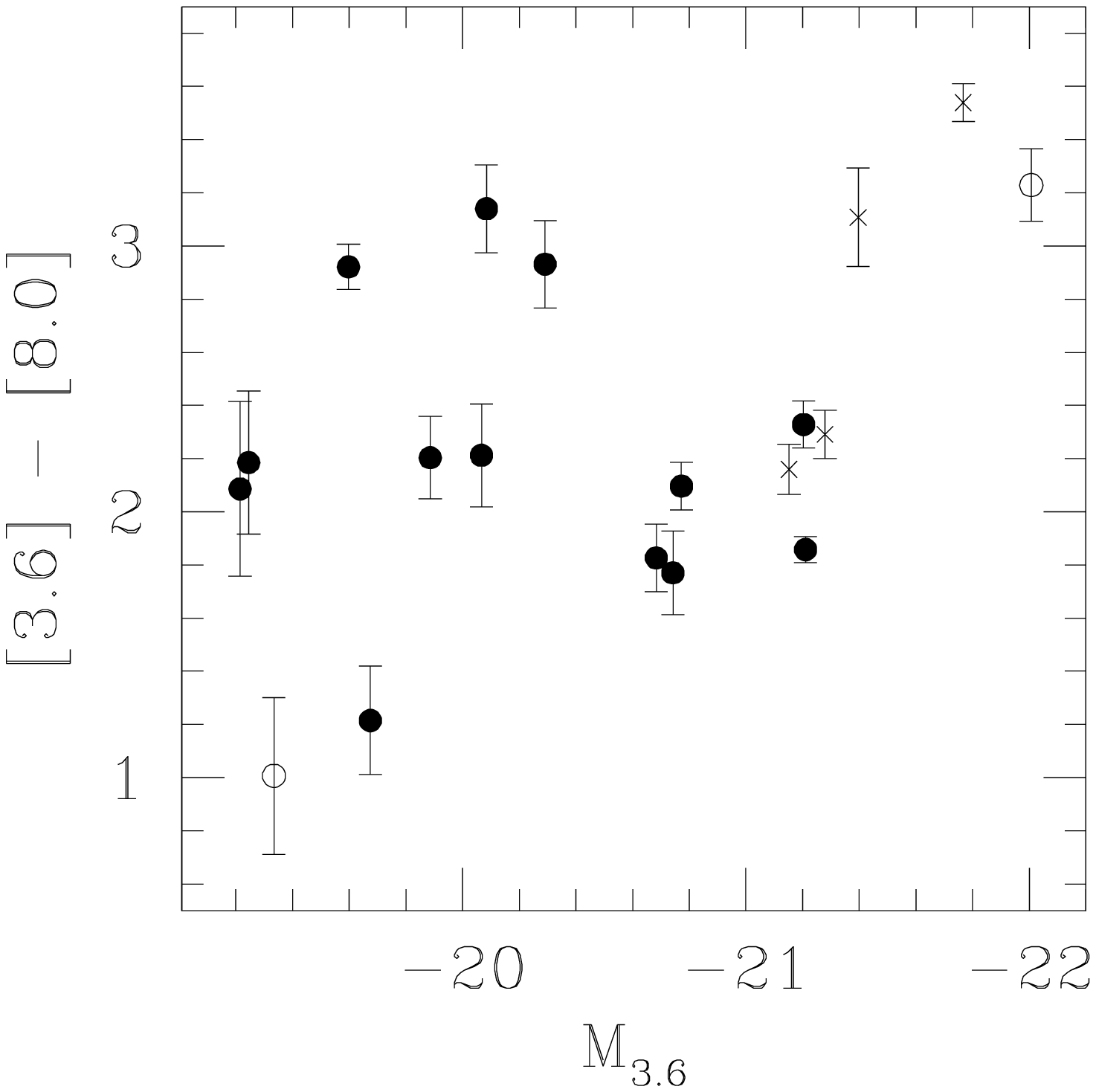}
\caption{The mid-infrared color of the KISS galaxies as a function of the 3.6
$\mu$m luminosity. The $\times$s represent the galaxies brighter than M$_B$ =
-18, the open circles
represent the galaxies for which no correction for internal extinction could be
estimated.}
\label{fig:masslum}
\end{figure}

The mid-infrared colors of the KISS galaxies are only slightly correlated with
the absolute 3.6 $\mu$m luminosities of the systems as shown in Figure
\ref{fig:masslum}. The figure indicates that the
bluest galaxies in the mid-infrared are the lowest luminosity systems, but the
colors have a spread of up to 2 magnitudes for any given 3.6 $\mu$m luminosity.

\subsection {Dust Properties}

Figure \ref{fig:colorplot} and all of the subsequent figures showing the [3.6]
$-$ [8.0] colors for the KISS galaxies indicate that these systems span a wide
range of mid-infrared colors. This range is an indication that some of the
galaxies have very high dust-to-stars ratios while others do not, as illustrated
in Figure \ref{fig:schematic}. The high dust-to-stars ratios are evidence for 
strong emission from hot dust and/or PAHs, but without spectroscopy it is
impossible to distinguish between a steep slope to the dust continuum and a PAH
emission line. However, the evidence from the study of low-metallicity galaxies 
by \citet{engelbracht2005} is that most low-metallicity systems have a deficit of
PAH emission, but an excess of hot dust. This analysis is certainly consistent
with the colors of these dwarf galaxies which also tend to be low-metallicity
systems.

\begin{figure}
\plotone{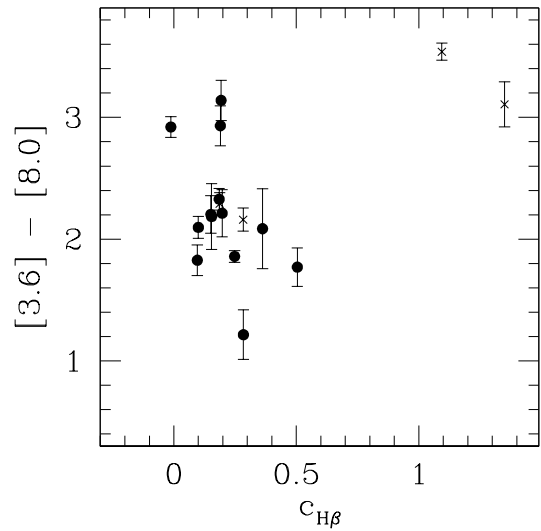}
\caption{The relationship between the [3.6] $-$ [8.0] color and the 
optical reddening at H$\beta$, c$_{H\beta}$. The filled circles are the true
dwarfs, the $\times$s are galaxies that are brighter than M$_B = -18$.}
\label{fig:chb}
\end{figure}

The reddening at H$\beta$ (c$_{H\beta}$) provides a line-of-sight measurement 
of the optical extinction from the dust \citep{osterbrock1989}. 
Figure \ref{fig:chb} shows the relationship between the [3.6] $-$ [8.0] 
color and the c$_{H\beta}$ ratio. Two of the most luminous galaxies in the
sample (brighter than the M$_B$=-18 cutoff after correction for extinction) have
the highest c$_{H\beta}$ ratios and very red mid-infrared colors. However, for
the rest of the galaxies in the sample, there is no correlation between the
c$_{H\beta}$ ratios and the mid-infrared color. The amount of line-of-sight dust 
absorption implied by the c$_{H\beta}$ parameter is not a good predictor of the 
amount of mid-infrared dust emission present in these galaxies.  The same result was 
found by Salzer \& MacAlpine (1988) for the FIR emission in galaxies.

\subsection {Metallicity}

\begin{figure}
\plotone{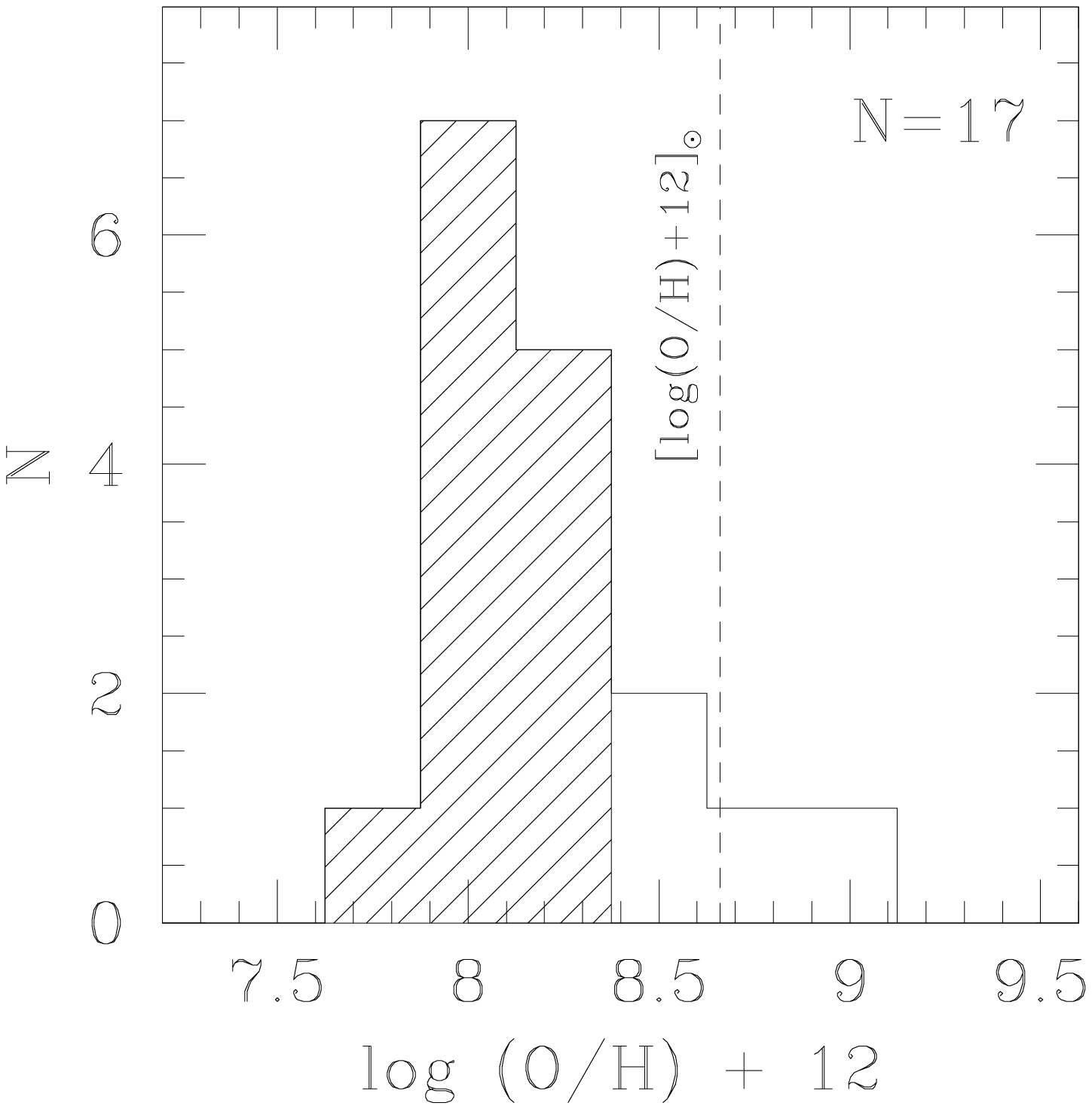}
\caption{The distribution of log(O/H)+12 for the KISS dwarf galaxies. The dashed
line indicates solar metallicity ([log(O/H)+12 = 8.66], \citealp{asplund2004}). 
While there are a couple of solar metallicity sources, all of them are luminous
objects. The shaded region shows the galaxies with M$_B$ fainter than -18 all of
which are significantly sub-solar with a median metallicity of 8.05 
(0.25 Z$_\odot$).}
\label{fig:abundhist}
\end{figure}

The KISS galaxies 
are low luminosity and mostly metal poor systems, making them potential analogs
of the galactic building blocks at high redshift. We use the emission-line
measurements from the follow-up spectroscopy to derive the [N II]/H$\alpha$ and
[O III]/H$\beta$ ratios for the galaxies. These ratios are used to determine the
galaxies' metallicity as described in \citet{salzer2005a}. Seventeen of the
nineteen galaxies have measured line ratios and, therefore, calculated 
metallicities. The other two galaxies possess noisy spectra where some of the
emission lines needed for deriving a metallicity estimate could not be measured.

While there is one super-solar metallicity source (KISS 2316), it has a
luminosity that is brighter than the M$_B$=18 cutoff after it is corrected for
extinction. All of the dwarf galaxies are significantly sub-solar 
([log(O/H)+12]$_\odot$ = 8.66, \citealp{asplund2004}). The four most metal-rich 
systems are more luminous than M$_B=-18$. If one ignores these higher luminosity
galaxies and considers only the dwarfs, then the median metallicity is 8.05 
(0.25 Z$_\odot$) and the galaxies span the range from 0.15 to 0.51 Z$_\odot$ as 
shown in Figure \ref{fig:abundhist}. In this figure the shaded histogram shows
the abundance distribution for the dwarfs while the open histogram shows the
distribution for the full sample.

\begin{figure*}
\plotone{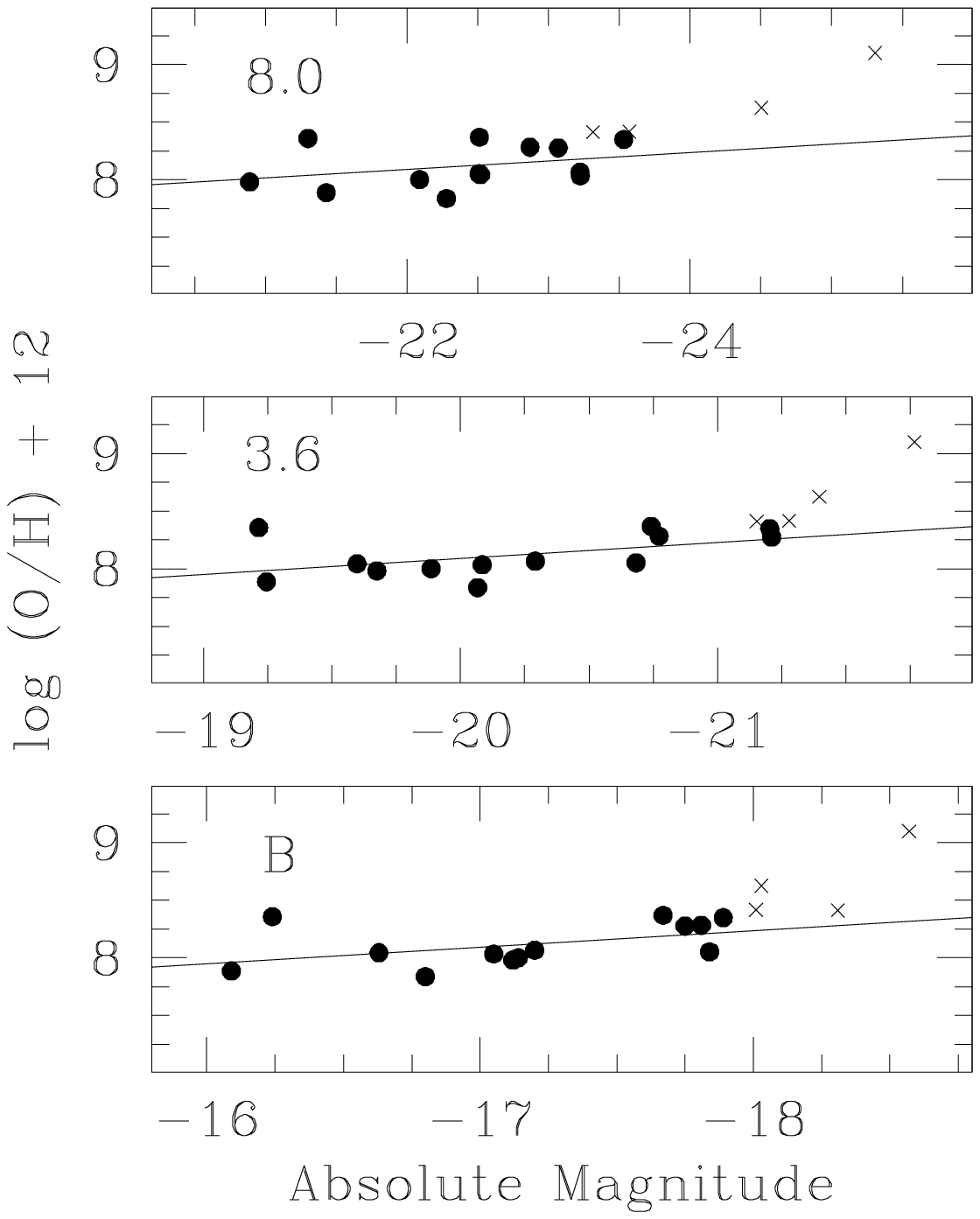}
\caption{The relationship between metallicity and luminosity measured in the B, 
3.6 $\mu$m, and 8.0 $\mu$m bands. The $\times$s indicate the galaxies brighter
than M$_B = -18$ after correcting for internal extinction. The lines indicate
linear least squares fits to the galaxies fainter than M$_B = -18$.}
\label{fig:lummet}
\end{figure*}

The correlation between luminosity and metallicity provides a probe of the 
relationship between the metallicity of a galaxy and either the stellar mass as 
measured by the $B$ or 3.6 $\mu$m luminosity or the mid-infrared emission from 
dust as measured by the 8.0 $\mu$m luminosity. Figure \ref{fig:lummet} shows these 
correlations for the KISS sample. The B-band magnitudes have been corrected for
extinction (for the 3 red, high c$_H\beta$ sources) -- the galaxies for which 
the internal extinction correction is uncertain do not have measured abundances
so they are not plotted. The fits to these relations include only the dwarf
galaxies in the sample. 

In addition to the correlation between the B-band and 3.6 $\mu$m
luminosities and the metallicity, the 8.0$\mu$m luminosity is also correlated with
metallicity. This result indicates that even in these low-metallicity systems
which might not have PAHs, the hot dust and/or PAH emission is correlated with
metallicity. 

The slope of the luminosity-metallicity relationship shown in Figure 
\ref{fig:lummet} is very shallow in all three bands -- -0.07 (8.0 $\mu$m), -0.16
(3.6 $\mu$m), and -0.16 (B-band). These slopes are in contrast with the slopes 
for a sample of KISS star-forming galaxies covering a large range in luminosity,
$-22 < $M$_B < -12$, which have a steeper B-band slope of -0.28 
(\citealp{salzer2005a}, computed using the \citep{edmunds1984} metallicity
calibration as was also done for the KISS galaxies presented here). However, 
the slope is consistent with those obtained for dwarf galaxy samples studied by
\citet{lee2004}, \citet{skillman1989} and \citet{richer1995} which have B-band slopes 
of -0.15 and by \citet{shi2005} whose dwarf sample has a slope of -0.17. These 
data provide further evidence for the dwarf galaxy relation having a flatter 
slope than that exhibited by the more luminous systems. The interpretation of 
the extremely shallow slope at 8.0 $\mu$m will have to wait until we have a 
better understanding of the nature of this emission (e.g., the exact mixture of
PAH and hot dust emission).

The $rms$ scatter in metallicity around these fits is 0.17 (8.0 $\mu$m), 
0.16 (3.6 $\mu$m), and 0.16 (B-band) respectively. If stellar mass and 
metallicity are correlated the scatter at 3.6$\mu$m should be tighter than in 
the B-band because extinction is less of a factor. We do not see any evidence 
for a lower scatter or a different slope at 3.6 $\mu$m.

\begin{figure}
\plotone{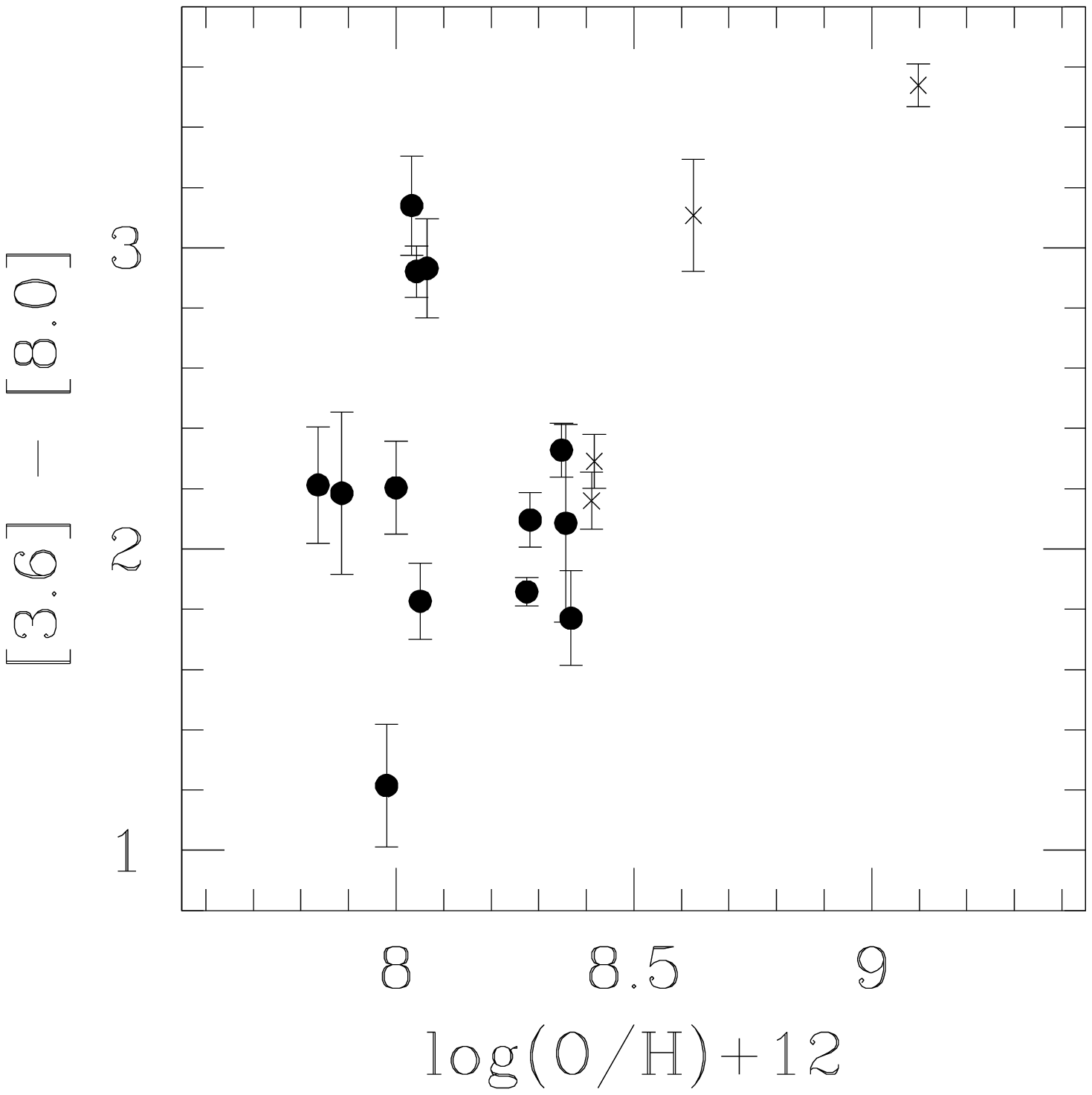}
\caption{The relationship between the [3.6] $-$ [8.0] $\mu$m color and metallicity.
The $\times$s indicates galaxies that are brighter than M$_B = -18$.}
\label{fig:colabund}
\end{figure}

Figure \ref{fig:colabund} shows the relationship between the mid-infrared color 
(dust-to-stars ratio) and the metallicity of the KISS galaxies. While the figure 
shows that the more metal-rich galaxies all have red mid-infrared colors,
three of the galaxies with the reddest [3.6] $-$ [8.0] are low-metallicity
systems. Clearly metallicity is not the primary factor driving the dust-to-stars
ratio in these galaxies, particularly for the low-luminosity systems. 

\subsection {Star-Formation Rates}

\begin{figure}
\plotone{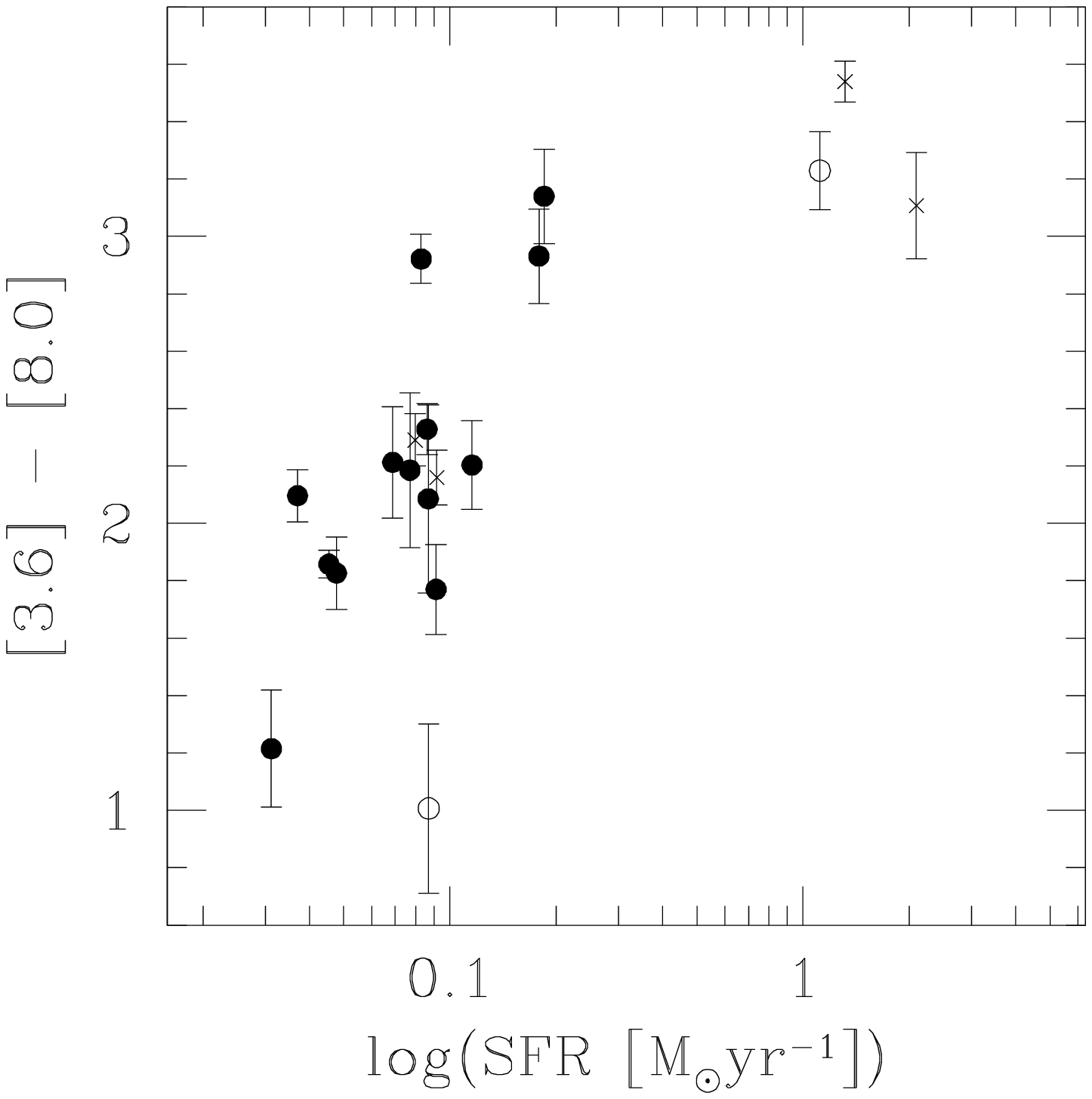}
\caption{The relationship between the [3.6] $-$ [8.0] $\mu$m color and the
star-formation rate in these galaxies as measured from the H$\alpha$ luminosity.
The filled circles are the true dwarfs, the $\times$s are galaxies that are 
brighter than M$_B = -18$, and the open circles are the galaxies
for which an internal extinction correction is not determined.}
\label{fig:sfrplot}
\end{figure}

The KISS dwarf galaxies are star-forming systems identified by the presence of
optical emission lines.  The distribution of star-formation rates in these
galaxies can be seen in Figure \ref{fig:sfrplot}. These star-formation
rates have been derived from the H$\alpha$ luminosity as in
\citet{kennicutt1998}, but the values have been corrected for the metallicity of
the systems. For the galaxies for which the metallicity is not determined (2
systems), the standard \citet{kennicutt1998} conversion is used. The metallicity 
correction takes into account the difference between the number of Lyman continuum
photons in solar metallicity models and models with lower metallicities that are 
closer in value to the observed abundances of our dwarf star-forming galaxies.  
The correction method is detailed in \citet{lee2002}. The radiation field in low 
metallicity systems is much harder so the inferred star-formation rate is
somewhat lower at a given H$\alpha$ luminosity compared to a metal-rich galaxy. 
Two of the three galaxies with high star-formation rates are luminous systems
while the third is one of the galaxies with a noisy spectrum so it might also be
a higher luminosity source. The star-formation rates for the dwarf galaxies are
modest for low-luminosity systems -- the rates are not as high as they are in
some of the most strongly star-forming blue compact dwarf galaxies but they are
still fairly strong for such low-luminosity dwarfs.

Figure \ref{fig:sfrplot} shows the relationship between mid-infrared color and
SFR for the KISS galaxies. The galaxies without an internal extinction correction 
do not have a measured metallicity so the SFR determined from the standard 
\citet{kennicutt1998} conversion is probably an over-estimate.
While Figure \ref{fig:colabund} shows that some of the low-metallicity systems in
this sample have red [3.6] $-$ [8.0], Figure \ref{fig:sfrplot} shows that
all of the red galaxies have high star-formation rates. The strength of the 8.0
$\mu$m dust emission is much more strongly correlated with the star-formation
rate than it is with the metallicity of these systems. 

\section {Discussion}

We have presented the observations of 19 star-forming galaxies from the KISS
survey observed with the IRAC camera on the Spitzer Space Telescope. Despite the
small fraction of star-forming dwarf galaxies detected with IRAS \citep{salzer1988}, 
we have detected all sample galaxies.  Many are quite bright at
mid-infrared wavelengths, as shown in Figures \ref{fig:fluxhist} and
\ref{fig:maghist}.  In particular, the emission detected 
at 8.0 $\mu$m indicates a significant amount of dust 
resides in some of these systems. 

One of the most striking features of the KISS sample in the mid-infrared is its
diversity. This is a complete sample of low-luminosity galaxies that show
evidence for star-formation through the presence of an H$\alpha$ emission line.
We find that the mid-infrared colors for some of the galaxies resemble those
of dust-poor, low star-formation early-type galaxies while others have colors as 
red or redder than the late-type galaxy population \citep{pahre2004}. The red 
colors are indicative of hot dust and/or PAH emission in these systems. The 
galaxies with bluer colors have low star formation rates and may also have a lower 
dust content or a lack of hot dust. If corruption of the 
3.6 $\mu$m flux by hot dust is a problem it might result in blue colors like
those seen for some of these dwarf galaxies. However, the shape of the SED shown
in Figure \ref{fig:schematic} and the correlation between
low star-formation rates and blue colors in these galaxies makes corruption of
the 3.6 $\mu$m flux by hot dust an unlikely explanation.

\citet{hogg2005} found that the low-luminosity galaxies in the overlap between 
the Sloan Digital Sky Survey and the Spitzer First Look Survey had significantly
bluer mid-infrared colors than the higher luminosity systems with similar
optical colors. The galaxies in the KISS sample also show a significant number
of systems with blue optical and mid-infrared colors. However, we also note that
in our sample and, in smaller numbers (possibly because they were not selecting
star-forming dwarf galaxies in particular) in the \citet{hogg2005} sample, the 
reddest low-luminosity galaxies are consistent with the reddest high
luminosity systems with comparable optical colors. In the \citet{hogg2005} sample
the reddest high luminosity objects have [3.6] $-$ [8.0] $\sim 3.9$ (all 
comparisons here are converted to the Vega magnitude system)  but most of the 
late-type galaxies have colors that are closer to [3.6] $-$ [8.0] $\sim
2.8$. The reddest low-luminosity galaxies
in the \citet{hogg2005} sample have [3.6] $-$ [8.0] $\sim 2.7$ while  the
reddest KISS galaxies have [3.6] $-$ [8.0] $\sim 3.1$.  

\citet{hogg2005} proposed that the deficit of red mid-infrared colors for 
low-luminosity galaxies in their sample could indicate that the supernova 
driven hot winds are driving dust and metals out the galaxies as happens to 
low-mass halos in the simulations of \citet{maclow1999}. The correlation between
luminosity and metallicity in galaxies has also been used as a probe of 
the relationship between a galaxy's stellar population and the build-up of 
metals. In the KISS sample we find a significant 
number of low-luminosity galaxies with very red mid-infrared colors which
clearly indicate the presence of hot dust and/or PAHs. This finding may indicate
that the dust is not escaping from these galaxies as has been conjectured or
that we are not probing the escaping dust. Instead  we may be probing a region 
very close to the star formation where the temperatures and possibly the dust 
properties in these low-metallicity systems might be different from those of 
the general galaxy population. This latter possibility is would help explain 
the observation that 8.0 $\mu$m dust emission is better correlated with
star-formation rate then it is with metallicity.

Figure \ref{fig:masslum} shows that some of the lowest mass galaxies, as 
indicated by their 3.6 $\mu$m luminosity, are also very blue and might be good 
candidates for dust expulsion by the winds. These are the same galaxies in
Figure \ref{fig:lummet} that are low luminosity at 3.6 $\mu$m and B-band and
are also low metallicity, possibly further corroborating the notion that some of
the metals and the dust is escaping. On the other hand, there are also low-mass 
galaxies that have red mid-infrared colors, indicating that they have been able 
to retain their dust, but are still low-metallicity systems. Clearly a 
measurement of the dynamical mass of these galaxies, rather than just a 
measurement of the stellar mass, is important 
for determining whether the galaxies with blue colors are the lowest mass halos.
Alternatively, these galaxies with blue colors in the mid-infrared may just be
producing less dust or be less efficient at heating it rather than losing it in 
winds since they have the lowest star-formation rates.

The correlation between B-band luminosity and metallicity shown in Figure
\ref{fig:lummet} has been observed before and our results are completely
consistent with the previous results for dwarf galaxies. It has been noted that
one would expect to see this relation extended into the mid-infrared since the
3.6 $\mu$m emission is also tied to the stellar mass of the galaxy and that is
indeed what this figure shows. The more surprising result is that we also see a
correlation between the 8.0 $\mu$m luminosity and metallicity indicating that
even in these low-metallicity systems, the emission from hot dust (or PAHs if 
they are present) is correlated with the metallicity of the system. It is
unclear whether this result is due to dust grain size, temperature distribution
in the star-forming regions, or some other physical property.
 
While we find that the 8.0 $\mu$m dust emission is correlated with star 
formation much more than it is with metallicity in the KISS galaxies (Figure
\ref{fig:sfrplot}), the connection between the hot dust/PAH
emission seen at these mid-infrared wavelengths and the dust responsible for
extinction at optical wavelengths is much less clear. For the 2 galaxies in our
sample with very high values of c$_{H\beta}$ (greater than 1), the mid-infrared
colors are very red indicating high dust-to-stars ratios. However, for
galaxies with lower values of c$_{H\beta}$, there is no evidence for a
correlation with the mid-infrared color. These results are consistent with IRAS
observations of star-forming dwarf galaxies that showed no 
correlation between c$_{H\beta}$ and far infrared luminosity 
\citep{salzer1988,kunth1985}. While the mid-infrared emission measured 
in these star-forming dwarf galaxies indicates that these systems contain hot
dust and/or PAHs, the amount of line-of-sight dust absorption implied by the 
c$_{H\beta}$ parameter is not a good predictor of the amount of mid-IR dust 
emission present as is also the case in the FIR \citep{salzer1988}. 

The mid-infrared observations of these star-forming dwarf galaxies have
highlighted some surprising features of these systems which need to be
investigated further. The spectral energy distributions from the
optical through 24 $\mu$m (and in a few cases through 160 $\mu$m) in these
galaxies will be presented in a forthcoming paper and will provide a much more
detailed look at the properties of the stars and dust. In
addition, it will be important to obtain systematic spectral observations of a
sample of these dwarf galaxies to resolve the question of whether the
mid-infrared emission is due to hot dust or the presence of a PAH feature at
these wavelengths.

\acknowledgments
This work is based on observations made with the Spitzer Space Telescope, which 
is operated by the Jet Propulsion Laboratory, California Institute of Technology 
under NASA contract 1407. Support for this work was provided by NASA through a
IRAC GTO award issued by JPL/Caltech under contract \# 1256790. JLR acknowledges 
support from NSF grant AST-0302049.  
JJS gratefully acknowledges support for the KISS project from the NSF through 
grants AST 95-53020, AST 00-71114, and AST 03-07766. This work also made use of 
data products provided by the NOAO Deep Wide-Field Survey (Jannuzi and Dey 1999; 
Jannuzi et al. 2005; Dey et al. 2005), which is supported by the National Optical 
Astronomy Observatory (NOAO). NOAO is operated by AURA, Inc., under a cooperative 
agreement with the National Science Foundation. Data products from the Two 
Micron All Sky Survey, which is a joint project of the University of Massachusetts 
and the Infrared Processing and Analysis Center/California Institute of Technology, 
funded by the National Aeronautics and Space Administration and the National 
Science Foundation were also used.

\clearpage
\begin{landscape}
\begin{deluxetable*}{cccccrccccrrrrrrr}
\tabletypesize{\tiny}
\tablewidth{8.7in}
\tablecaption{Kitt Peak International Spectroscopic Survey Data \label{tab:KISS}}
\tablehead{
\colhead{KISS\#} & \colhead{Field} & \colhead{ID} & \colhead{RA} & \colhead{Dec} & 
\colhead{Vel.} & \colhead{B} & \colhead{M$_B$} & \colhead{M$_{B0}$
\tablenotemark{1}} & \colhead{(B-V)$_0$ \tablenotemark{2}} &
\colhead{c$_{H\beta}$ \tablenotemark{3}} & \colhead{EW$_{H\alpha}$} & 
\colhead{log L$_{H\alpha}$} & \colhead{[NII]/H$\alpha$} & \colhead{[OIII]/H$\beta$} & 
\colhead{Abund. \tablenotemark{4}} & \colhead{SFR \tablenotemark{5}}\\
\colhead{} & \colhead{} & \colhead{} & \colhead{J2000} & \colhead{J2000} & 
\colhead{km s$^{-1}$} & \colhead{} & \colhead{} & \colhead{} & \colhead{} & \colhead{} &
\colhead{$\AA$} & \colhead{ergs s$^{-1}$} & \colhead{} & \colhead{} & \colhead{} &
\colhead{M\solar\ yr$^{-1}$}}
\startdata
 2292 & H1426 &  8731  &  14:25:09.2  & 35:25:15.9  &  8659 & 17.62 & -17.80 & -18.31 & 0.66  &  0.283&  38.4  &  40.39 & -0.756 &  0.398 &  8.411 & 0.09 \\
 2300 & G1426 &  6944  &  14:26:08.9  & 33:54:19.8  & 10271 & 19.70 & -16.09 & -16.09 & 0.50  &  0.154& 301.6  &  40.52 & -1.566 &  0.676 &  7.886 & 0.08 \\
 2302 & H1426 &  5703  &  14:26:17.5  & 35:21:35.5  &  8342 & 18.18 & -17.16 & -17.16 & 0.42  & \nodata &  25.1  &  40.04 & \nodata & \nodata & \nodata & 0.09 \\
 2309 & G1426 &  4670  &  14:26:53.6  & 34:04:14.5  &  7231 & 17.89 & -17.12 & -17.12 & 0.43  &  0.284&  81.8  &  40.09 & -1.402 &  0.653 &  7.980 & 0.03 \\
 2316 & G1426 &  1167  &  14:28:14.9  & 33:30:25.7  & 10685 & 18.51 & -17.38 & -18.57 & 0.81  &  1.094&  32.2  &  41.01 & -0.475 & -0.662 &  9.098 & 1.32 \\
 2318 & H1426 &   299  &  14:28:24.6  & 35:10:21.5  & 22163 & 19.66 & -17.88 & -17.88 & 0.95  & \nodata &  22.5  &  41.15 & \nodata & \nodata & \nodata & 1.12 \\
 2322 & G1430 &  9529  &  14:29:09.6  & 32:51:26.7  &  8574 & 17.53 & -17.84 & -17.84 & 0.53  &  0.096&  73.2  &  40.24 & -1.289 &  0.415 &  8.051 & 0.05 \\
 2326 & G1430 &  7761  &  14:29:32.7  & 33:30:40.3  &  7935 & 18.07 & -17.14 & -17.14 & 0.51  &  0.151& 189.8  &  40.65 & -1.370 &  0.662 &  8.000 & 0.11 \\
 2338 & H1430 &  4786  &  14:30:27.9  & 35:32:07.2  & 11689 & 18.89 & -17.20 & -17.20 & 0.77  &  0.190& 258.7  &  40.81 & -1.266 &  0.747 &  8.065 & 0.18 \\
 2344 & H1430 &  3113  &  14:31:03.6  & 35:31:14.8  &  4166 & 16.06 & -17.75 & -17.75 & 0.44  &  0.248& 217.2  &  40.12 & -1.041 &  0.468 &  8.275 & 0.04 \\
 2346 & G1430 &  3224  &  14:31:14.4  & 33:19:13.2  & 10819 & 19.09 & -16.80 & -16.80 & 0.67  &  0.197& 122.8  &  40.49 & -1.661 &  0.800 &  7.836 & 0.07 \\
 2349 & H1430 &  3139  &  14:31:20.0  & 34:38:03.8  &  4396 & 17.30 & -16.63 & -16.63 & 0.52  & -0.012& 396.4  &  40.48 & -1.300 &  0.665 &  8.043 & 0.08 \\
 2357 & G1430 &  1974  &  14:31:39.2  & 33:26:32.3  & 10759 & 18.21 & -17.67 & -17.67 & 0.58  &  0.504&  27.4  &  40.40 & -0.808 &  0.461 &  8.368 & 0.09 \\
 2359 & H1430 &  1039  &  14:31:49.3  & 35:28:40.0  & 22512 & 20.02 & -17.55 & -18.03 & 0.64  &  1.350&  76.7  &  41.66 & -0.452 &  0.161 &  8.625 & 2.09 \\
 2368 & G1430 &   376  &  14:32:18.9  & 33:02:53.7  & 10972 & 18.88 & -17.05 & -17.05 & 0.76  &  0.193& 245.4  &  40.84 & -1.316 &  0.818 &  8.033 & 0.18 \\
 2382 & H1434 &  5378  &  14:34:08.0  & 34:19:34.4  &  6813 & 17.07 & -17.81 & -17.81 & 0.35  &  0.100&  33.4  &  40.03 & -1.049 &  0.440 &  8.282 & 0.04 \\
 2398 & H1437 &  8675  &  14:36:33.1  & 34:58:04.4  &  9006 & 17.50 & -18.01 & -18.01 & 0.46  &  0.187&  35.8  &  40.33 & -0.764 &  0.368 &  8.417 & 0.08 \\
 2403 & G1437 &  5761  &  14:37:42.6  & 33:36:26.7  & 12047 & 19.91 & -16.24 & -16.24 & 0.36  &  0.362& 103.8  &  40.38 & -0.798 &  0.506 &  8.357 & 0.09 \\
 2406 & H1437 &  3649  &  14:38:27.8  & 35:08:59.0  &  8641 & 17.51 & -17.89 & -17.89 & 0.49  &  0.185&  19.4  &  40.38 & -0.915 &  0.394 &  8.348 & 0.09 \\
\tablenotetext{1}{B-band absolute magnitude corrected for internal extinction using the ad hoc 
method from \citet{melbourne2002} (This correction is only applied to the three
galaxies in the sample that are optically red and have a high c$_{H\beta}$. For
all other galaxies M$_{B0} = $ M$_B$.)}
\tablenotetext{2}{B-V color corrected for Galactic absorption}
\tablenotetext{3}{decimal reddening coefficient}
\tablenotetext{4}{log (O/H) + 12, which is a measure of metallicity. For details of the 
calculation see \citet{melbourne2002} and \citet{salzer2005a}.}
\tablenotetext{5}{star-formation rate calculated from the H$\alpha$ luminosity 
using the prescription of \citet{kennicutt1998} and then corrected for the 
effect of metallicity when an abundance measurement is available using the 
prescription of \citet{lee2002}. Two of the galaxies in the sample do not 
have measured abundances because the lines in these spectra are not high 
enough signal-to-noise to determine robust values.}
\enddata													 
\end{deluxetable*}
\clearpage												 
\end{landscape}

\begin{deluxetable*}{crrrrrrrrr}
\tabletypesize{\small}
\tablewidth{6.8in}
\tablecaption{2MASS and NOAO Wide Deep Survey Photometry}
\tablehead{
\colhead{KISS\#} & \colhead{Field} & \colhead{ID} & \colhead{J$_{2MASS}$} & 
\colhead{H$_{2MASS}$} & \colhead{K$_{2MASS}$} & \colhead{B$_{NDWFS}$} & 
\colhead{R$_{NDWFS}$} & \colhead{I$_{NDWFS}$} & \colhead{K$_{NDWFS}$}}
\startdata
 2292 & H1426 &  8731 & \nodata &\nodata &\nodata &  17.67&  16.50 & 16.03  &14.55 \\
 2300 & G1426 &  6944 & \nodata &\nodata &\nodata &  19.76&  19.12 & 18.81  &17.53 \\
 2302 & H1426 &  5703 & \nodata &\nodata &\nodata &  18.20&  17.43 & 17.07  &14.89 \\
 2309 & G1426 &  4670 & \nodata &\nodata &\nodata &  17.93&  17.16 & 16.73  &15.75 \\
 2316 & G1426 &  1167 &  16.0 & 15.2 & 14.8   &  18.50&  17.11 & 16.57  &\nodata \\
 2318 & H1426 &   299 & \nodata &\nodata &\nodata & 19.58&  18.11 & 17.46  & 15.77 \\
 2322 & G1430 &  9529 & \nodata &\nodata &\nodata & 17.56&  16.66 & 16.21  & \nodata \\
 2326 & G1430 &  7761 & \nodata &\nodata &\nodata & 18.03&  17.35 & 17.06  & \nodata \\
 2338 & H1430 &  4786 & \nodata &\nodata &\nodata & 18.94&  18.06 & 17.73  & 16.01 \\
 2344 & H1430 &  3113 & \nodata &\nodata &\nodata & 17.14&  16.50 & 16.22  & 15.22 \\
 2346 & G1430 &  3224 & \nodata &\nodata &\nodata & 19.07&  17.97 & 17.30  & \nodata \\
 2349 & H1430 &  3139 &  16.3 & 16.2 & 15.5  &  17.33&  16.58 & 16.32  &14.90 \\
 2357 & G1430 &  1974 & \nodata &\nodata &\nodata &  18.21&  17.23 & 16.85  &\nodata \\
 2359 & H1430 &  1039 & \nodata &\nodata &\nodata &  20.07&  18.88 & 18.38  &16.50 \\
 2368 & G1430 &   376 & \nodata &\nodata &\nodata &  18.87&  18.02 & 17.72  &\nodata \\
 2382 & H1434 &  5378 &  16.4 & 15.0 & 15.5  &  17.11&  16.37 & 15.99  &14.51 \\
 2398 & H1437 &  8675 & \nodata &\nodata &\nodata &  17.60&  16.68 & 16.26  &14.82 \\
 2403 & G1437 &  5761 & \nodata &\nodata &\nodata &  19.99&  19.16 & 18.81  &\nodata \\
 2406 & H1437 &  3649 &  16.0 & 15.4 & 15.2  &  17.57&  16.63 & 16.20  &\nodata \\
\enddata
\end{deluxetable*}

\begin{deluxetable*}{ccccccccccccc}
\tabletypesize{\small}
\tablewidth{6.8in}
\tablecaption{Spitzer IRAC Photometry \label{tab:IRAC}}
\tablehead{
\colhead{KISS\#} & \colhead{Field} & \colhead{ID} & \colhead{R$_{3.6}$ [\arcs]
\tablenotemark{1}} & \colhead{Ellip. \tablenotemark{2}} & \colhead{[3.6]} & 
\colhead{$\sigma_{3.6}$} & 
\colhead{[4.5]} & \colhead{$\sigma_{4.5}$} & \colhead{[5.8]} & 
\colhead{$\sigma_{5.8}$} & \colhead{[8.0]} & \colhead{$\sigma_{8.0}$}}
\startdata
2292 & H1426 &  8731 & 12 & 0.11 & 14.31 & 0.07 & 14.29 & 0.09 & 13.53 & 0.17 & 12.15 & 0.07 \\
2300 & G1426 &  6944 & 10 & 0.32 & 16.59 & 0.19 & 16.32 & 0.24 & 15.96 & 0.51 & 14.40 & 0.19 \\
2302 & H1426 &  5703 & 14 & 0.32 & 16.05 & 0.15 & 16.03 & 0.21 & 15.81 & 0.47 & 15.04 & 0.25 \\
2309 & G1426 &  4670 & 12 & 0.16 & 15.40 & 0.11 & 15.43 & 0.16 & 14.92 & 0.32 & 14.18 & 0.17 \\
2316 & G1426 &  1167 & 14 & 0.07 & 14.15 & 0.06 & 14.09 & 0.08 & 12.61 & 0.11 & 10.61 & 0.03 \\
2318 & H1426 &   299 & 14 & 0.35 & 15.50 & 0.11 & 15.31 & 0.15 & 14.68 & 0.28 & 12.27 & 0.07 \\
2322 & G1430 &  9529 & 20 & 0.22 & 14.76 & 0.08 & 14.68 & 0.11 & 14.31 & 0.24 & 12.93 & 0.10 \\
2326 & G1430 &  7761 & 14 & 0.19 & 15.39 & 0.11 & 15.33 & 0.15 & 14.56 & 0.27 & 13.18 & 0.11 \\
2338 & H1430 &  4786 & 14 & 0.18 & 15.82 & 0.13 & 15.49 & 0.16 & 14.60 & 0.27 & 12.89 & 0.09 \\
2344 & H1430 &  3113 & 70 & 0.70 & 12.66 & 0.03 & 12.53 & 0.04 & 11.56 & 0.07 & 10.80 & 0.04 \\
2346 & G1430 &  3224 & 12 & 0.08 & 15.88 & 0.14 & 15.80 & 0.19 & 15.25 & 0.37 & 13.67 & 0.14 \\
2349 & H1430 &  3139 & 20 & 0.32 & 14.39 & 0.07 & 14.10 & 0.09 & 12.94 & 0.13 & 11.47 & 0.05 \\
2357 & G1430 &  1974 & 20 & 0.57 & 15.19 & 0.10 & 15.01 & 0.13 & 14.89 & 0.31 & 13.42 & 0.12 \\
2359 & H1430 &  1039 & 12 & 0.05 & 16.14 & 0.16 & 15.96 & 0.20 & 15.69 & 0.45 & 13.03 & 0.10 \\
2368 & G1430 &   376 & 14 & 0.37 & 15.89 & 0.14 & 15.45 & 0.16 & 14.57 & 0.27 & 12.75 & 0.09 \\
2382 & H1434 &  5378 & 20 & 0.37 & 14.17 & 0.06 & 14.16 & 0.09 & 13.47 & 0.16 & 12.07 & 0.06 \\
2398 & H1437 &  8675 & 30 & 0.44 & 14.27 & 0.07 & 14.17 & 0.09 & 13.39 & 0.16 & 11.98 & 0.06 \\
2403 & G1437 &  5761 & 14 & 0.35 & 16.96 & 0.23 & 16.78 & 0.29 & 15.81 & 0.47 & 14.88 & 0.23 \\
2406 & H1437 &  3649 & 20 & 0.14 & 14.25 & 0.06 & 14.26 & 0.09 & 13.40 & 0.16 & 11.93 & 0.06 \\
\tablenotetext{1}{The semi-major axis of the aperture in which the flux was
measured}
\tablenotetext{2}{The ellipticity of the fit aperture}
\enddata
\end{deluxetable*}

\end{document}